\documentclass{PoS}

\title{Mixed action simulations: approaching physical quark masses}

\ShortTitle{Approaching physical quark masses}

\author{S.~D\"urr$^{1}$, Z.~Fodor$^{1,2,3}$, C.~Hoelbling$^{2,5}$, S.D.~Katz$^{2,3}$, \speaker{S.~Krieg}$^{,2,4}$, Th.~Kurth$^2$, L.~Lellouch$^{5}$, Th.~Lippert$^{1,2,4}$, K.K.~Szabo$^2$, G.~Vulvert$^5$ \\ $^1$John von Neumann Institute for Computing (NIC), DESY, D-15738, Zeuthen $/$ FZJ, D-52425, Juelich, Germany \\ $^2$Department of Physics, University of Wuppertal, D-42097 Wuppertal, Germany \\ $^3$Institute for Theoretical Physics, E\"otv\"os University, H-1117 Budapest, Hungary \\ $^4$J\"ulich Supercomputing Center (JSC), Forschungszentrum J\"ulich, D-52425 J\"ulich, Germany \\ $^5$Centre de Physique Th\'eorique, Case 907, Campus de Luminy,
    F-13288 Marseille Cedex 9, France\thanks{CPT is ``UMR 6207 du CNRS
    et des universit\'es d'Aix-Marseille I, d'Aix-Marseille II et du
    Sud Toulon-Var, affili\'ee \`a la FRUMAM.''}

\\
E-mail: \email{s.krieg@fz-juelich.de}}       

\abstract{Some algorithmic details of our $N_f=2+1$ QCD mixed action
simulations with overlap valence and improved Wilson sea quarks are presented.
}

\FullConference{The XXV International Symposium on Lattice Field Theory\\
                July 30 - August 4 2007\\
                Regensburg, Germany}

\begin{document}


\section{Motivation}


Although they have desirable theoretical properties, Ginsparg-Wilson (GW)
\cite{GinspargWilson} fermion formulations are both computationally and
conceptually demanding when it comes to simulations including effects of
dynamical quarks (see eg.\ \cite{Fodor:2003bh, Cundy:2005pi, paper2, paper3}).
One way to circumvent this problem is the mixed action approach: the ensemble
production is performed with a relatively cheap fermion discretization and GW
fermions are only used for calculations in the valence sector.
This allows to speed up the simulation while retaining most advantages of the
GW fermions, such as their exact chiral symmetry and the resulting
absence of complicated operator mixings in valence calculations.
However, because valence and sea quarks are different, the
theory suffers from unitarity violations. As a result, the chiral
perturbation theory
formulae needed to extrapolate lattice data in quark mass and lattice spacing become more
involved and have more free parameters.


\section{Action and algorithm details}


\subsection{Choice of action}

To generate the configurations we use the L\"uscher-Weisz gauge action
\cite{LuescherWeisz} and stout-smeared \cite{StoutLinks} $O(a)$-improved
\cite{CloverFermions} Wilson fermions, where the improvement coefficient
$c_\mathrm{SW}$ is taken at tree level.
The combination of link-fattening and $O(a)$ improvement greatly reduces chiral
symmetry breaking effects \cite{FatClover}.
This action has an improved scaling behavior where the theoretically leading
$O(\alpha_s a)$ contributions will, in practice, be negligible and the scaling
to the continuum looks as if the theory had only $O(a^2)$ cut-off effects 
\cite{FatClover, Hoffmann:2007nm}.
We use 2, 3 and 6 levels of stouting and several $\beta$-values in order to
check the scaling behavior explicitly.
The smearing will also significantly reduce the appearance of small eigenvalues
related to short distance artifacts.
This will improve the convergence rate of the solvers, as discussed in section
\ref{sect:HMC}.
In the valence sector we use overlap fermions \cite{OverlapFermions} with ``UV filtering''
 to improve the locality without altering $\rho=1$
\cite{FatOverlap}.


\subsection{Choice of parameters}

{\it Setting the strange sea quark mass:}
The approximate determination of the strange mass is done by $N_f=3$
simulations: at a given $\beta$ we search for the quark mass where the relation
\begin{equation}
m_{PS}/m_V=\sqrt{2m_{K}^2-m_{\pi}^2}/m_{\phi} 
\end{equation}
is satisfied.
We have determined the $\beta$ dependence of this approximate strange mass in a
fairly large range ($\beta=2.9-3.8$) and it turns out to be smooth.


{\it Matching sea and valence quarks:}
The different discretizations in the
sea and valence sectors
lead to discretization error induced unitarity violating effects. As
far as low-energy properties are concerned, these effects, as well as
those associated with any mismatch between sea and valence quark masses,
can be accounted for with the appropriate version
of mixed action partially quenched chiral perturbation theory (MAPQXPT),
as described in the
accompanying proceedings contribution by L. Lellouch \cite{Laurent}. Accordingly, a
precise matching of the
sea and valence sector is not needed. Still, in order to remain
relatively close to the unitary situation, we have one set of valence
data where the pion and kaon masses
of the two sectors are approximately matched.


{\it Overview of simulation points:}
A brief summary of our ensembles is given in Tab.\,\ref{tb:spect}.

\begin{table}
\begin{center}
\begin{tabular}{|c|ccc|c|c|}
\hline
$\beta$& $a$\,[fm] &  $m_\pi$\,[MeV]  & $L^3\cdot T$ & overlap inversion \\
\hline
3.3  & 0.136 & 360 & $16^3\cdot 64$ & DONE \\
     &       & 310 & $24^3\cdot 64$ & DONE \\
     &       & 250 & $24^3\cdot 64$ & DONE \\
\hline
3.57 & 0.088 & 570 & $24^3\cdot 64$ & DONE \\
     &       & 490 & $24^3\cdot 64$ & DONE \\
     &       & 410 & $24^3\cdot 64$ & DONE \\
     &       & 300 & $32^3\cdot 64$ & DONE \\
     &       & 190 & $48^3\cdot 64$ & DONE \\
\hline
3.7  & 0.069 & 520 & $32^3\cdot 96$ & DONE \\
     &       & 400 & $32^3\cdot 96$ & RUNNING \\
     &       & 290 & $40^3\cdot 96$ & DONE \\
\hline
\end{tabular}
\caption{\label{tb:spect} $N_f=2+1$ simulation points. The last column
indicates the status of the overlap inversions.}
\end{center}
\end{table}


\subsection{Dynamical fermion algorithm\label{sect:HMC}}

We aim to run simulations with 2+1 flavors, with pion masses approaching the
physical point.
To simulate the two light flavours we use the Hybrid Monte Carlo (HMC)
\cite{HMC} algorithm with even/odd preconditioned \cite{EvenOdd} clover
fermions.
However, in the regime of light quark masses the standard HMC suffers from
``critical slowing down'', that is on top of the increased computational cost
per trajectory, the autocorrelation times grow significantly.
Several improvements over the standard version have been proposed, many of
which can be combined.
We use the following ones:
\begin{itemize}
\vspace*{-2mm}
\itemsep-2pt
\item
multiple time-scale integration (``Sexton-Weingarten integration scheme'')
\cite{SextonWeingarten} to be able to run the computationally most demanding
part of the simulation (the inversion of the light fermion matrix) at a larger
time-step then the comparatively less costly part,
\item
mass preconditioning (``Hasenbusch trick'') \cite{Hasenbusch}, to reduce jumps
in the fermionic force and
\item
Omelyan integrator \cite{OmelyanIntegrator} to reduce the energy violations
during the Molecular Dynamics (MD) part of the HMC.
\vspace*{-2mm}
\end{itemize}
The strange quark is included via the RHMC algorithm.
This method is exact and highly efficient when
combined with the Sexton-Weingarten integration scheme \cite{RHMC}. 


\subsection{Mixed precision solver}

The most time consuming part, both in valence and see sector calculations, is
the (incomplete) fermion matrix inversion by means of a solver.
In order to maintain reversibility, the MD part of the HMC algorithm has to be
performed in double precision.
The same holds true for propagator calculations at small quark masses, due to
the large condition number of the fermion matrix involved.
However, this does not imply that each fermion matrix multiplication needs to
be done in double precision.
In the valence sector we wish to solve
\begin{equation}
Dx = b
\label{eqn:propagator}
\end{equation}
with $D$ the overlap or clover operator to construct the correlator.
In the sea sector we wish to solve
\begin{equation}
D^\dagger Dx = b
\label{eqn:determinant}
\end{equation}
for the clover action to calculate the fermionic force within the MD.
To accelerate the solvers it is, in either case, possible to use a single
precision version of $D$ within a mixed precision solver.
We find that there is basically no penalty in terms of the iteration count;
the increase of the number of forward multiplications is well below 10\%.

\begin{figure}
\includegraphics[width=15cm]{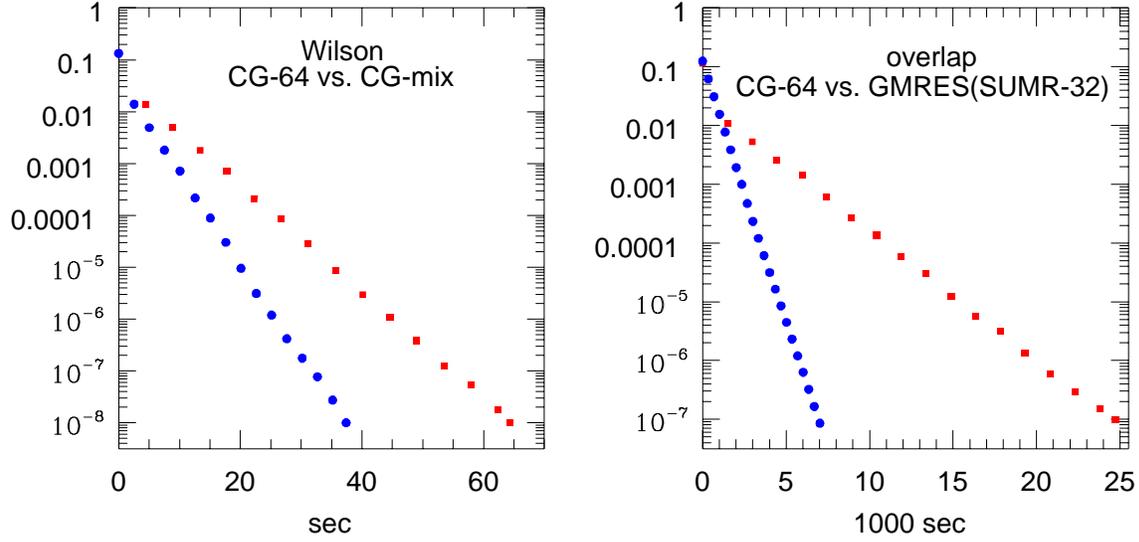}
\vspace*{-2mm}
\caption{Performance of CG-64 (double precision, red squares) and CG-mix (mixed
precision, blue circles) during the fat clover MD trajectory (left).
Chirally projected CG (double precision) versus relaxed GMRESR(SUMR-32)
(mixed precision) during overlap propagator calculations (right).\label{fig:cg}}
\end{figure}

\begin{figure}
\includegraphics[width=7.5cm]{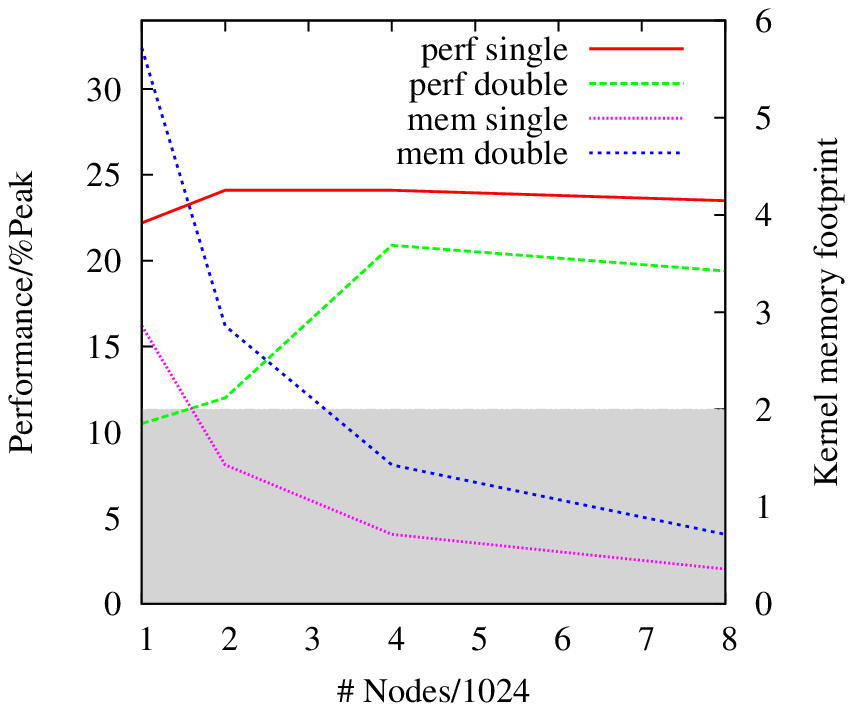} 
\includegraphics[width=7.5cm]{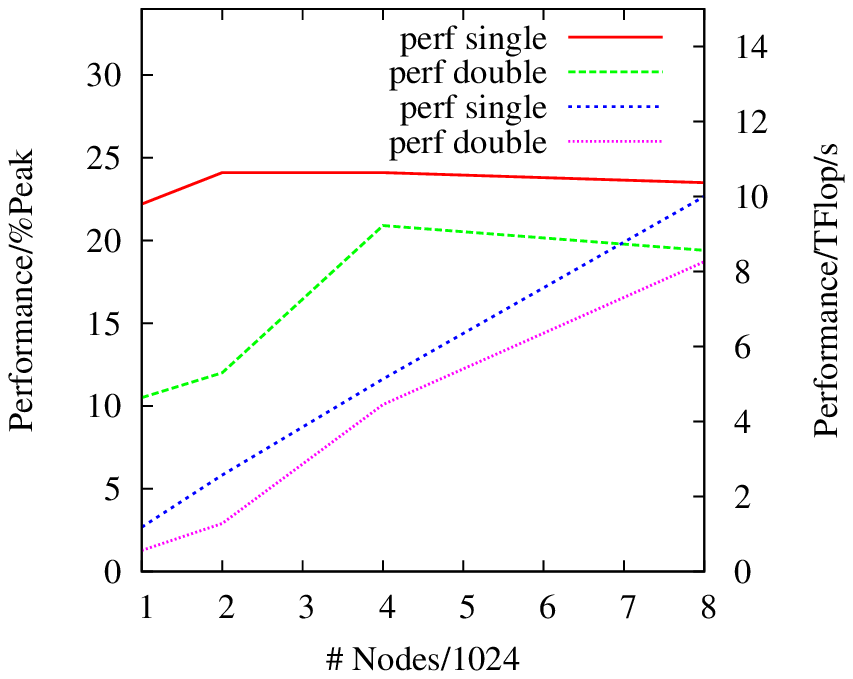} 
\vspace*{-8mm}
\caption{Strong scaling analysis of the Wilson fermion matrix multiplication on
a global $48^3\times 96$ lattice. Left: performance of the single and double
precision kernel in percentage of machine peak, as well as their memory
footprint for 1024 up to 8192 nodes (2048 to 16384 CPUs). The gray shaded area
indicates the size of the L3 cache. Right: the same scaling analysis and kernel
performance in TFlops.\label{fig:kernel}}
\end{figure}

How a single precision calculation can be used to accelerate a solver is most
transparent in the valence sector where we use the relaxed GMRESR~\cite{paper3}
algorithm with a recursive SUMR~\cite{paper2} preconditioning.
In this scheme, the SUMR is merely used to calculate a low precision inversion.
Therefore, the SUMR can be coded in single precision (except for global sums),
even if the GMRESR requires
double precision accuracy.
Since almost all matrix multiplications are performed within the SUMR, the
whole solver is dominated by the single precision matrix multiplication
performance, resulting in a significant speedup (see Fig.\,\ref{fig:cg}).
The gain is due to three effects:
\begin{itemize}
\vspace*{-2mm}
\itemsep-2pt
\item
On a generic computer architecture
the peak performance of the single precision solver will be
larger than that of the double precision version.
\item
The performance of the solver is usually bound by the bandwidth to the system
memory.
Thus, on the same architecture, usually twice as many single precision than
double precision numbers can be loaded from system memory per unit time.
\item
The single precision matrix vector multiplication routine requires half the
memory of the double precision version.
The inverter will fit into the cache for larger local lattice sizes and
the range in which the algorithm scales with the number of nodes will improve
(see Fig.\,\ref{fig:kernel}).
\vspace*{-2mm}
\end{itemize}


\subsection{Phase structure and algorithm stability}

\begin{figure}
\begin{center}
\hspace*{-12mm}\includegraphics[width=17cm]{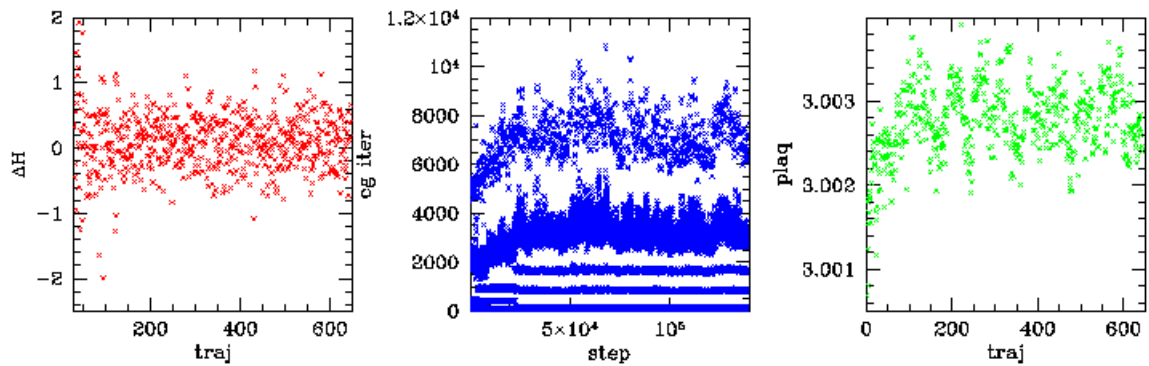}
\end{center} 
\vspace*{-10mm}
\caption{$\Delta H$ (left), CG count (center) and plaquette (right) for the
fat clover action at $m_\pi=190$\,MeV.\label{fig:most2}}
\end{figure}

During the HMC evolution a number of observables is monitored to detect any
potential instability of the algorithm.
In Fig.\,\ref{fig:most2} the energy violation $\Delta H$, the CG iteration
number needed to reach the residue tolerance and the plaquette are shown for
the run with 190\,MeV pion mass. After the thermalization a stable distribution can be seen.
\begin{figure}
\includegraphics[width=15cm]{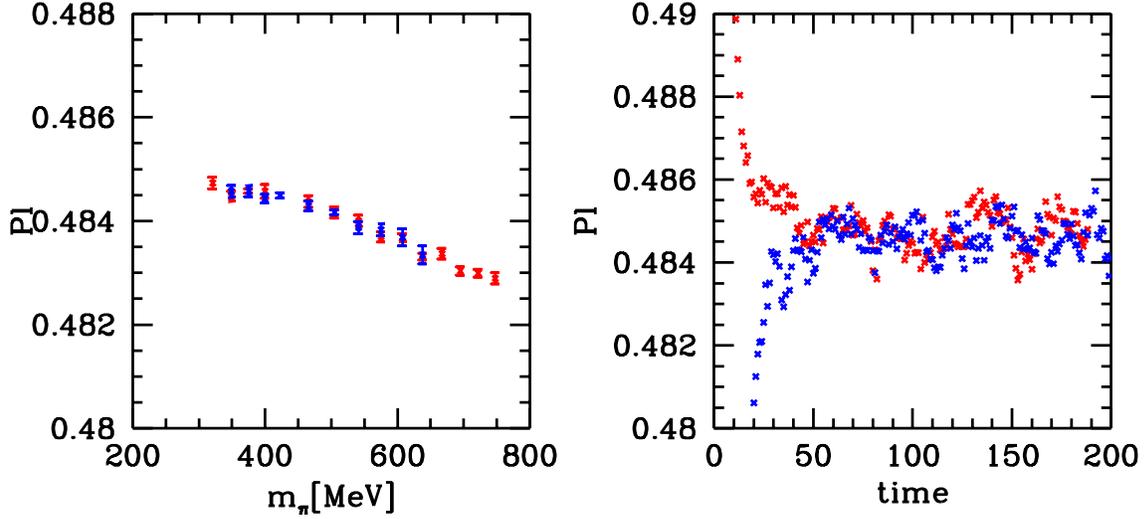}
\vspace*{-2mm}
\caption{Thermalized plaquette versus $m_\pi$ (left) and cold versus hot start
plaquette at $\beta=3.5$ (right).\label{fig:phase}}
\end{figure}
When attempting a ``thermal cycle'' in the pion mass one finds no signs of
a hysteresis.
Moreover, ``cold'' and ``hot'' starts quickly lead to the same plaquette
(see Fig.\,\ref{fig:phase}).
Altogether, it seems that we are far from any potential bulk phase transition.


\subsection{Overall performance and strong scaling analysis \label{sect:perf}}

\begin{figure}
\hfill
\includegraphics[width=7.4cm]{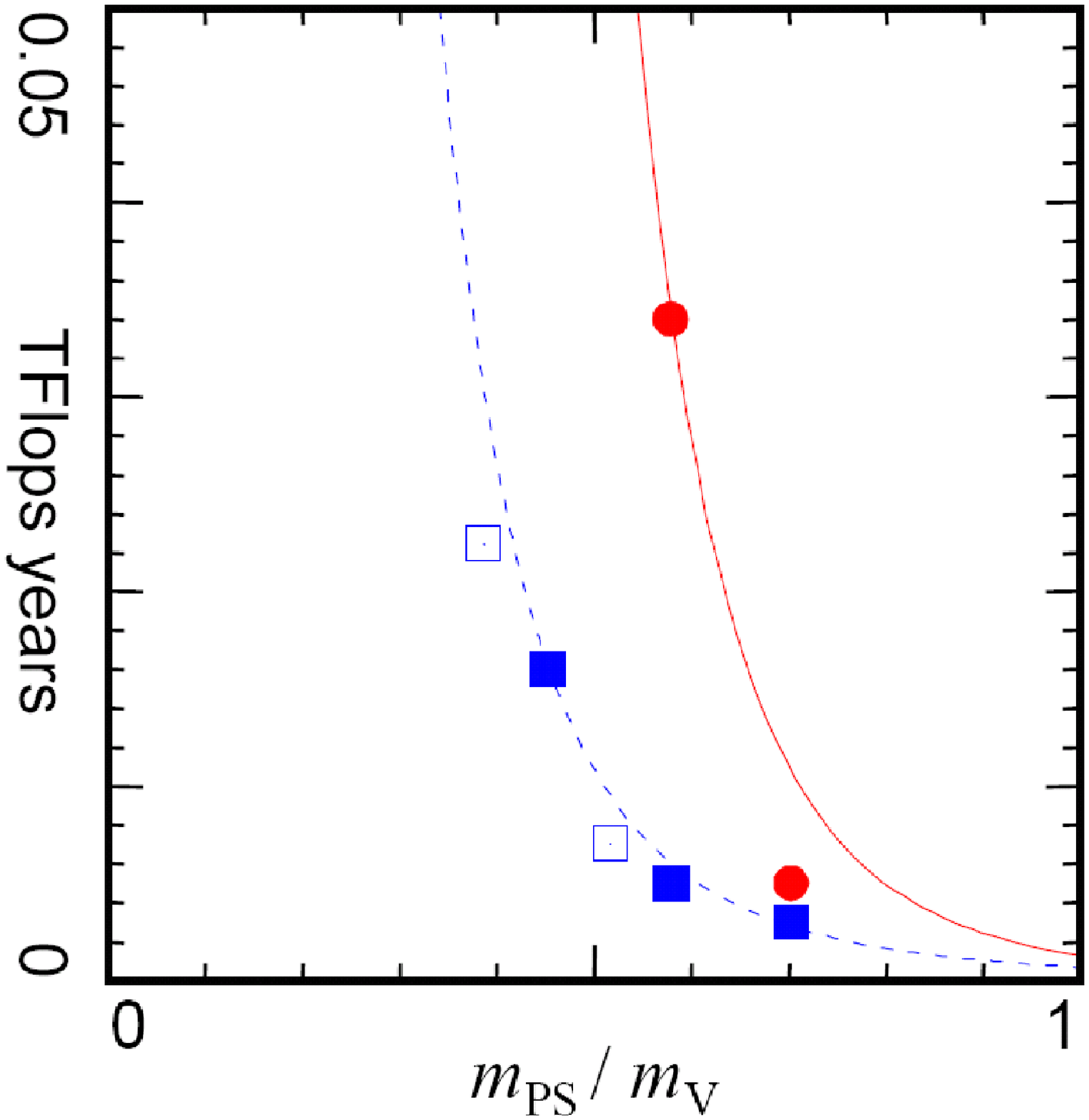}
\includegraphics[width=7.2cm]{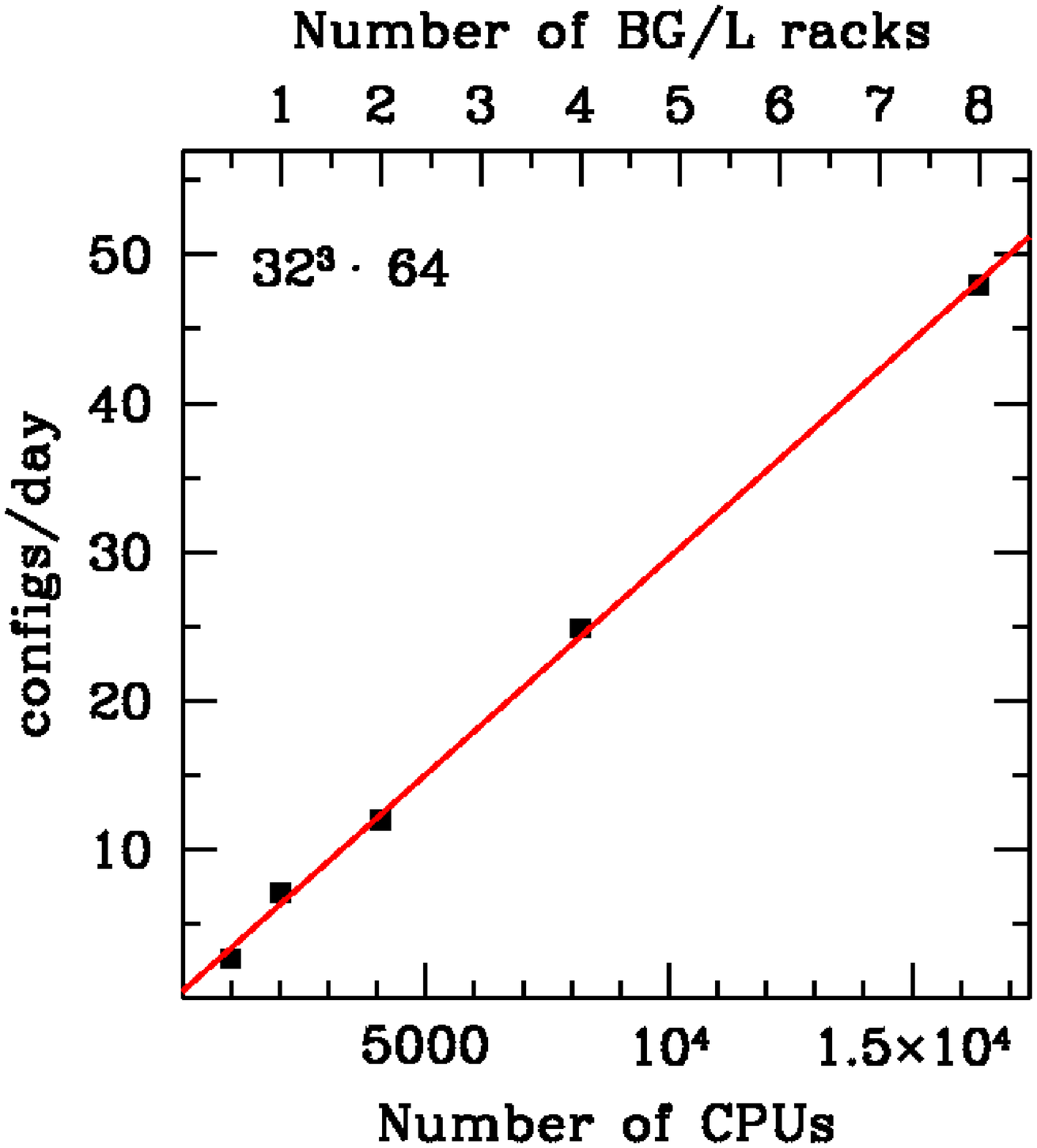}
\vspace*{-2mm}
\caption{Left: updated ``Berlin Wall'' plot -- full circles correspond to 
the pre-2005 dynamical Wilson simulations, solid squares represent the 
algorithm of \cite{JansenAlg}, open squares are this work. Right: Scaling 
of our algorithm with the number of CPUs. \label{Fig:Performance}}
\end{figure}

Several improved versions of the HMC algorithm can be found in the literature.
As shown in the left panel of Fig.\,\ref{Fig:Performance}, our
algorithm is compatible with the performance reported in
\cite{LuescherAlg,JansenAlg}.

The right panel shows the good scaling properties of our HMC variety with the
number of CPUs of the Blue Gene/L of the J\"ulich Supercomputing Centre (JSC).
Within errors, the curve is perfectly linear, very much as for the Wilson
kernel shown in Fig~\ref{fig:kernel}.
The bottom line is that we can simulate dynamical quarks in a regime well
below 250\,MeV.


\section{Outlook \label{sect:results}}

\begin{figure}
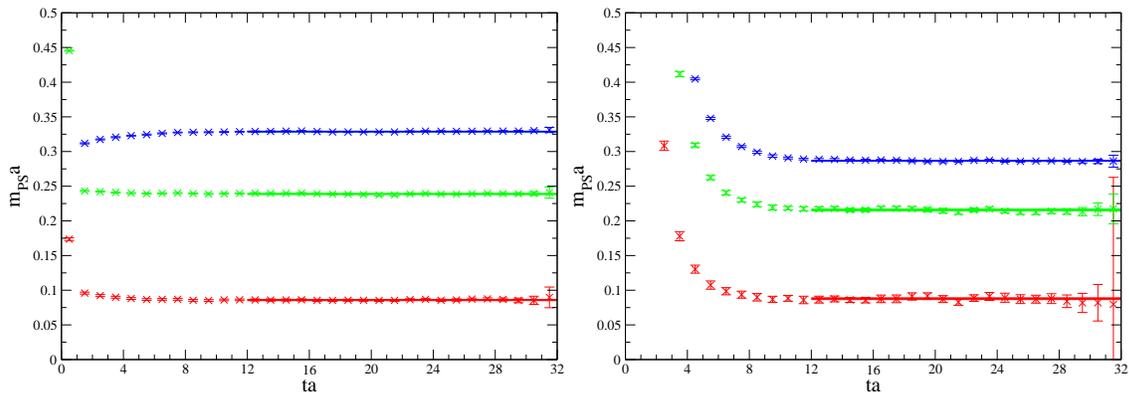

\includegraphics[width=7.4cm]{wpi_0483plat_mult.eps}
\includegraphics[width=7.4cm]{ovpi_0483plat.eps}
\vspace{-2mm}
\caption{Mass plateaus for clover (left) and overlap (right) pions, kaons and
etas at $m_\pi=190$\,MeV. \label{fig:pions}}
\end{figure}

To illustrate the statistical quality of our results, Fig.~\ref{fig:pions} presents
effective mass plateaus of charged pions, kaons and non-singlet $s\bar
s$ pseudoscalar mesons, for our lightest sea pion at $\beta=3.57$ (cf.
Tab.~\ref{tb:spect}) -- both with clover (left) and overlap (right) valence quarks.

The performance and stability of our simulations, the statistical
accuracy of our results, as well as our growing understanding of their
chiral behavior \cite{Laurent} are very promising and we look forward to
presenting phenomenological results, with controlled extrapolations to
the physical point of $2{+}1$ flavor QCD, in future publications.

%

\vspace{0.4cm}
\noindent
{\bf Acknowledgments}\newline
The numerical work is performed on the IBM BlueGene/Ls at FZ J\"ulich and on PC clusters at the University of Wuppertal and at CPT Marseille. This work is supported in part by EU grant I3HP, OTKA grant AT049652, DFG grant FO 502/1, EU RTN contract MRTN-CT-2006-035482 (FLAVIAnet) and by the CNRS's GDR grant n$^\mathrm{o}$2921 (``Physique subatomique et calculs sur r\'eseau'').




\begin{thebibliography}{99}


\enlargethispage{2mm}
\vspace*{-2mm}
\itemsep-2pt

\bibitem{GinspargWilson}
P.~H.~Ginsparg and K.~G.~Wilson,
Phys.\ Rev.\ D {\bf 25} (1982) 2649.

\bibitem{Fodor:2003bh}
   Z.~Fodor, S.~D.~Katz and K.~K.~Szabo,
   JHEP {\bf 0408} (2004) 003 [hep-lat/0311010].

\bibitem{Cundy:2005pi}
  N.~Cundy, S.~Krieg, G.~Arnold, A.~Frommer, T.~Lippert and K.~Schilling,
  hep-lat/0502007.

\bibitem{paper2}
  G.~Arnold {\it et al.}
  , hep-lat/0311025.

\bibitem{paper3}
  N.~Cundy, J.~van den Eshof, A.~Frommer, S.~Krieg, T.~Lippert and
  K.~Sch\"afer,
  Comput.\ Phys.\ Commun.\  {\bf 165} (2005) 221 [hep-lat/0405003].

\bibitem{LuescherWeisz}
M.~L\"uscher, P.~Weisz,
Phys.\ Lett.\ {\bf B158} (1985) 250.

\bibitem{StoutLinks}
C.~Morningstar, M.~Peardon,
Phys.\ Rev.\ {\bf D69} (2004) 054501 [hep-lat/0311018].

\bibitem{CloverFermions}
B.~Sheikholeslami and R.~Wohlert,
Nucl.\ Phys.\ B {\bf 259} (1985) 572.

\bibitem{FatClover}
  T.~A.~DeGrand, A.~Hasenfratz and T.~G.~Kovacs [MILC Collaboration],
  hep-lat/9807002.
  C.~W.~Bernard and T.~A.~DeGrand,
  Nucl.\ Phys.\ Proc.\ Suppl.\  {\bf 83} (2000) 845 [hep-lat/9909083].
  S.~Capitani, S.~D\"urr and C.~Hoelbling,
  JHEP {\bf 0611} (2006) 028 [hep-lat/0607006].

\bibitem{Hoffmann:2007nm}
   R.~Hoffmann, A.~Hasenfratz and S.~Schaefer,
   0710.0471 [hep-lat].

\bibitem{OverlapFermions}
	H.~Neuberger,
	Phys.\ Lett.\ B {\bf 417} (1998) 141 [hep-lat/9707022].


\bibitem{FatOverlap}
  S.~D\"urr, C.~Hoelbling and U.~Wenger,
  JHEP {\bf 0509} (2005) 030 [hep-lat/0506027].

\bibitem{Laurent}
L.~Lellouch {\it et al.},
PoS(LAT2007) 115.

\bibitem{HMC}
S.~Duane, A.~D.~Kennedy, B.~J.~Pendleton, D.~Roweth,
Phys.\ Lett.\ {\bf B195} (1987) 216.

\bibitem{EvenOdd}
T.~A.~DeGrand and P.~Rossi,
Comput.\ Phys.\ Commun.\ {\bf 60} (1990) 211.

\bibitem{SextonWeingarten}
J.~C.~Sexton, D.~H.~Weingarten,
Nucl.\ Phys.\ {\bf B380} (1992) 665.

\bibitem{Hasenbusch}
M.~Hasenbusch,
Phys.\ Lett.\ {\bf B519} (2001) 177 [hep-lat/0107019].

\bibitem{OmelyanIntegrator}
T.~Takaishi, P.~de~Forcrand,
Phys.Rev. {\bf E73} (2006) 036706 [hep-lat/0505020].

\bibitem{RHMC}
  M.~A.~Clark, B.~Joo and A.~D.~Kennedy,
  Nucl.\ Phys.\ Proc.\ Suppl.\  {\bf 119} (2003) 1015 [hep-lat/0209035].
  M.~A.~Clark and A.~D.~Kennedy,
  Phys.\ Rev.\ Lett.\ {\bf 98} (2007) 051601 [hep-lat/0608015].

\bibitem{LuescherAlg}
  M.~L\"uscher,
  Comput.\ Phys.\ Commun.\  {\bf 165} (2005) 199 [hep-lat/0409106].

\bibitem{JansenAlg}
  C.~Urbach, K.~Jansen, A.~Shindler and U.~Wenger,
  Comput.\ Phys.\ Commun.\ {\bf 174} (2006) 87 [hep-lat/0506011].

\end{thebibliography}
\end{document}